\documentclass[aps,pre,11pt,longbibliography]{revtex4-1}
\usepackage{amssymb}
\usepackage{bbold}
\usepackage{mathtools}
\usepackage{bm}
\usepackage{graphicx}
\usepackage{color}

\begin{document}
\title{Condensate in quasi two-dimensional turbulence}
\author{S. Musacchio}
\affiliation{Department of Physics, University of Torino,  via P. Giuria 1, 10125 Torino, Italy}
\author{G. Boffetta}
\affiliation{Department of Physics and INFN, University of Torino,  via P. Giuria 1, 10125 Torino, Italy}
\begin{abstract}
We investigate the process of formation of large-scale structures in a
turbulent flow confined in a thin layer.
By means of direct numerical simulations of the Navier-Stokes
equations, forced at an intermediate scale $L_f$,
we obtain a split of the energy cascade in which one fraction 
of the input goes to small scales generating the three-dimensional 
direct cascade. The remaining energy flows to large scales 
producing the inverse cascade which eventually causes the formation of a 
quasi two-dimensional condensed state at the largest horizontal scale.
Our results shows that the connection between the two actors of the split
energy cascade in thin layers is tighter than what was established before:
the small scale three-dimensional turbulence acts as an effective viscosity
and dissipates the large-scale energy thus providing a viscosity-independent
mechanism for arresting the growth of the condensate.
This scenario is supported by quantitative predictions of the 
saturation energy in the condensate.
\end{abstract}
\maketitle 
In many instances, geophysical and astrophysical 
flows are confined in thin layers of small aspect ratio 
either by material boundaries or by other physical mechanisms which constrain the motion. 
The thickness of such layers can be much smaller than
the typical horizontal scales, while being at the same time much
larger than the dissipative viscous scales.
Turbulent flows in such quasi two-dimensional (2D) geometries display
an interesting phenomenology with both 2D and three-dimensional (3D) features. 
Numerical \cite{smith1996crossover,celani2010turbulence,musacchio2017split,alexakis2018cascades} 
and experimental \cite{shats2010turbulence,byrne2011robust,xia2011upscale,xia2017two} 
works have demonstrated the emergence of a {\it split energy cascade} 
in which a fraction of the energy flow to large scales (as in a pure 2D flow)
and the remaining part goes to small scales producing the 3D direct cascade.
The key parameter which controls the relative flux of the two energy
cascades is the geometric ratio $S=L_z/L_f$ between the confining scale $L_z$ 
and the forcing scale $L_f$ \cite{smith1996crossover,celani2010turbulence}.
In particular, it has been shown that there exists a critical ratio
$S^*$ above which the inverse cascade is suppressed and the thin layer
recovers the usual 3D phenomenology~\cite{celani2010turbulence,benavides2017critical}.
In the limit $S \to 0$, when the thickness becomes smaller than the 
viscous scale, vertical motion is suppressed and the flow fully
recovers the 2D phenomenology.

The bidimensionalization of the flow, and in particular the value of
$S^*$, is affected by other physical factors, besides confinement. 
Rotation along the confined direction $z$
in general favors the bidimensionalization, increasing
the relative intensity of the inverse flux 
at given $S$ with respect to
the non-rotating case \cite{smith1999transfer,lindborg2005effect,
deusebio2014dimensional,pouquet2017dual}. 
Conversely, a stable stratification of the
density produces an increase of the effective dimensionality of the flow
and suppresses the large-scale energy transfer 
\cite{brethouwer2007scaling,sozza2015dimensional}. 

The inverse energy cascade generates a very long, 
non-stationary transient with an increasing value of kinetic energy
of the flow. A fraction of the energy injected at scale
$L_f$ goes to the large scales where it is not dissipated by viscosity
and, for finite horizontal extensions, accumulates producing a
large-scale vortex system called the {\it condensate} 
\cite{hossain1983long,smith1993bose}. 
The statistics of the condensate has been investigated in details by
experiments \cite{sommeria1986experimental,xia2008turbulence,xia2009spectrally, xia2017two} 
and numerical simulations 
\cite{chertkov2007dynamics,gallet2013a,laurie2014universal,frishman2017turbulence,frishman2017jets} 
in the $2D$ limit $S=0$. 
In the case of a square box with periodic boundary conditions,
the condensate is a pair of system-size vortices of opposite sign. The
vorticity profile of these vortices has been shown to displays universal
features, independent of the forcing mechanism which produces the 
inverse cascade 
\cite{chertkov2007dynamics,laurie2014universal}.
Changing the shape of the domain from square to rectangular, 
the emergence of jets in the condensed state with a complex phenomenology
has also been observed \cite{bouchet2009random,frishman2017jets}.  

The growth of the energy of the condensate can be arrested by 
different mechanisms.
The presence of a linear friction force $\alpha {\bm u}$ (as in
\cite{laurie2014universal,frishman2017turbulence,frishman2017jets}) 
causes the saturation of the energy to the value 
$E_c \simeq \varepsilon_{inv}/2\alpha$, where $\varepsilon_{inv}$ is
the flux of the inverse cascade. 
Further, in the case in which the forcing is correlated in time, it
has been shown that the fast sweeping due to the large scale velocity
decorrelates the forcing and the velocity field. 
As a consequence, the energy input rate vanishes at long times causing
the saturation of the condensate~\cite{gallet2013a}. 
Even in the absence of a friction force and in the ideal case 
of a forcing which guarantees a constant energy input 
(as in the case of random-in-time forcing), any finite viscosity 
will eventually produce a sink of energy at large
scales, thus arresting the growth of the condensate at finite energy.
In this case, the value of the asymptotic energy $E_c$ is determined 
by the balance
between the flux of the inverse cascade $\varepsilon_{inv}$ 
and the viscous dissipation at the scale of the condensate $L$, 
$\varepsilon_{inv} \simeq 2 \nu E_c / L^2$, 
which gives the dimensional predictions
$E_c \sim \varepsilon_{inv} L^2 /\nu$~\cite{eyink1996exact}. 
Estimating the time required to reach the steady state as  
$t_c \simeq E_c/\varepsilon_{inv}$ one gets $t_c \sim L^2 /\nu$~\cite{eyink1996exact}. 
In the limit of very large Reynolds number $Re$, equivalent to vanishing 
viscosity (which is relevant for geophysical applications), 
the flux of the inverse cascade becomes independent 
on the small scale viscosity~\cite{boffetta2012two}, and therefore 
the asymptotic energy of the condensate grows without limits
(i.e. $E_c, t_c \to \infty$ as $\nu \to 0$).

In this paper we show that this divergence is removed when a thin
layer, with finite $S$, is considered. The direct energy
cascade at scales below $L_f$ produces a small-scale 3D flow which
acts as an effective, eddy viscosity 
which, in the large $Re$ limit, becomes independent of the value 
of the molecular viscosity.  
This eddy viscosity arrests the condensate at an energy level 
much smaller than that obtained by 
the dimensional estimate $E_c \sim \varepsilon_{inv} L^2 /\nu$. 

To investigate quantitatively this prediction, we
performed a set of direct numerical simulations of the 3D Navier-Stokes
equation for an incompressible velocity field ${\bm u}({\bm x},t)$ 
(with ${\bm \nabla} \cdot {\bm u}=0$) 
\begin{equation}
{\partial {\bm u} \over \partial t} + {\bm u} \cdot {\bm \nabla} {\bm u} =
- {\bm \nabla} p + \nu \nabla^2 {\bm u} + {\bm f}
\label{eq1}
\end{equation}
where the constant density has been adsorbed into the pressure $p$ and
 $\nu$ is the kinematic viscosity.
The two-dimensional forcing ${\bm f}$ is restricted to the two 
horizontal components (2D2C) ${\bm f}({\bm x})=(f_x(x,y),f_y(x,y),0)$.
It is Gaussian, white in time and in Fourier space is confined 
in a narrow cylindrical shell of wavenumbers centered around
$K_f=2 \pi/L_f$. 
Thanks to the delta-correlation in time, the rates of injection of 
energy $\varepsilon$ and of enstrophy 
$\beta= K_f^2 \varepsilon$ do not depend on the flow, 
and they are kept fixed. 
Simulations are performed in a triply periodic domain 
with horizontal sizes $L_x=L_y = 2 \pi$ 
and aspect ratio $r=L_x/L_z$, 
by means of a fully parallel, fully dealiased 
pseudospectral code with a second-order Runge-Kutta time stepping and 
explicit integration of the linear part. 
The resolution is $N_x=N_y=r N_z=1024$
with uniform grid for two aspect ratios $r=32$ and $r=64$. 
The wavenumber of the forcing is fixed at $K_f=8$.
The characteristic time and kinetic energy 
at the forcing scale are defined as
$t_f = \beta^{-1/3}$ and $E_f = \varepsilon t_f$. 

Simulations of the split cascade in a thin layer are very 
demanding numerically since their need to resolve phenomena at 
very different scales: the horizontal box (of size $L_x=L_y$) must
be much larger than the forcing scale $L_f$ for the development of
the inverse cascade which also requires $L_f>L_z$ 
\cite{celani2010turbulence}. Finally, scale separation between 
$L_z$ and the viscous scale $\eta=(\varepsilon^3/\nu)^{1/4}$ is needed
to produce the direct energy cascade.
Therefore, to increase the extension of the direct inertial range, 
the viscous term in (\ref{eq1}) is replaced 
by an hyperviscous term $(-1)^{p-1}\nu_p \nabla^{2p} {\bm u}$ with $p=8$ 
and $\nu_p=10^{-37}$.
We do not use any large-scale dissipation (such as linear friction). 

\begin{figure}[h]
\includegraphics[width=0.49\columnwidth]{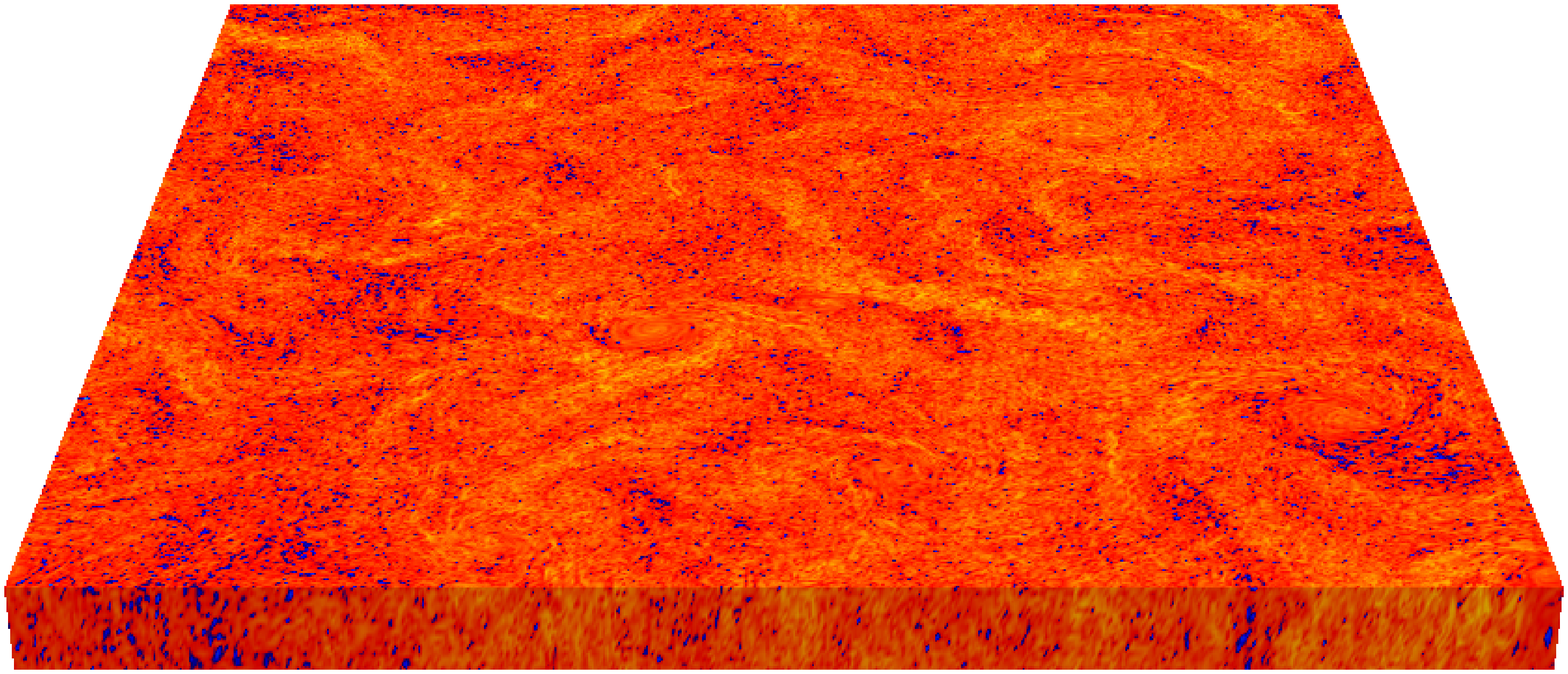}
\includegraphics[width=0.49\columnwidth]{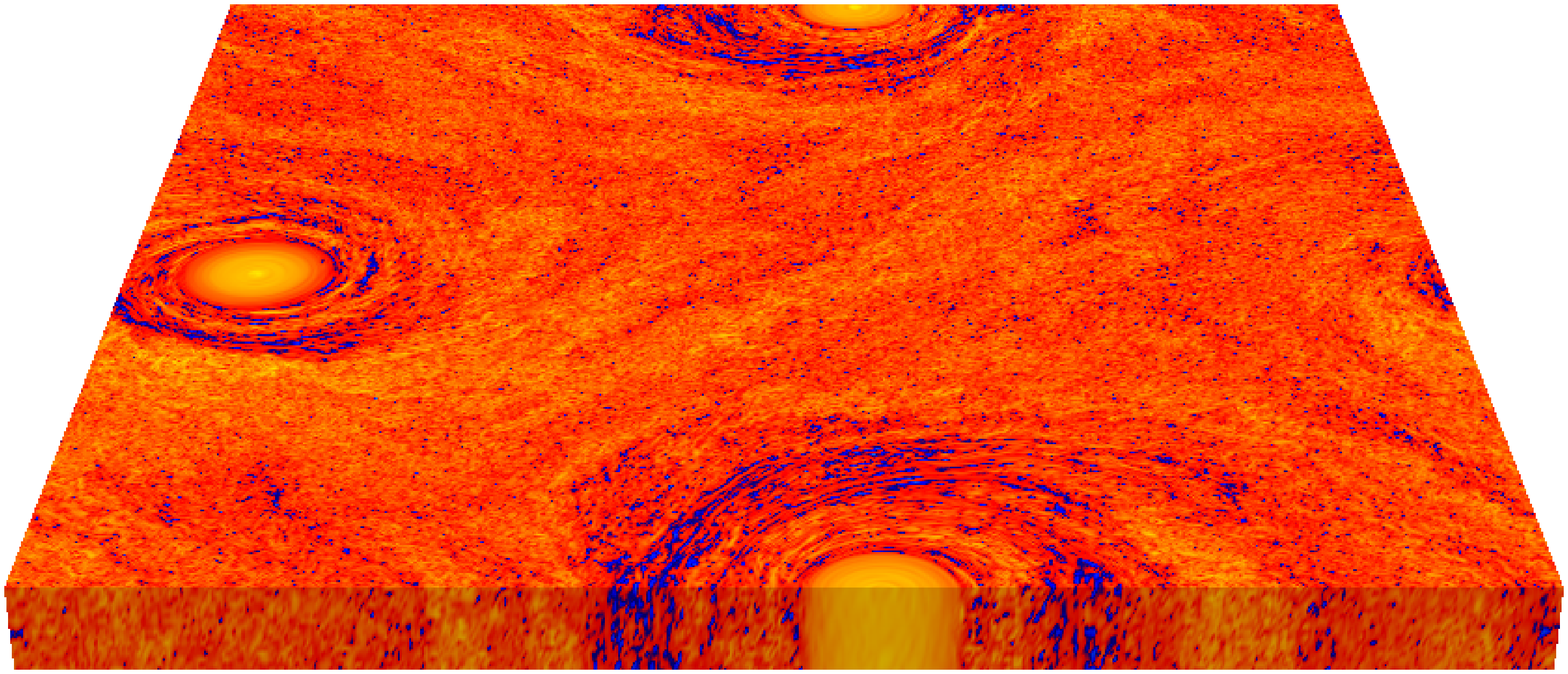}
\caption{Snapshots of the square vertical vorticity for the simulation
at $S=1/4$ at times $t=24 t_f$ (left panel) and $t=1200 t_f$ (right panel). 
The same logarithmic color scale is used for the two cases, with 
blue/yellow representing small/large values.
For clarity, the vertical scale has been stretched by a factor $2$.}
\label{fig1}
\end{figure}

At $t=0$, the velocity field is initialized to zero 
plus a small random perturbation, which triggers the 3D instability. 
The energy of the initial perturbation is $E_{pert} \simeq 1.4 \cdot
10^{-7}  E_f$. The total simulation time is very long, in order
to allow a complete development of the condensed state. 
In Figure~\ref{fig1} we show 
two snapshots of the vertical component of the
vorticity field ${\bm \omega} = \nabla \times {\bm u}$ 
at intermediate times and
in the late stage of the simulation for $S=1/4$. 
During the first stage of the evolution ($t=24t_f$, left panel) 
the vorticity field is characterized by small-scale structures, 
with some organization induced by the large-scale velocity produced by the
inverse cascade.
The condensate becomes clearly visible at late times ($t=1200 t_f$, 
right panel).
The vorticity field is dominated by a quasi-two-dimensional dipole, 
surrounded by small-scale three-dimensional turbulence. We observe
that 3D structures are observable also inside the vortex structure.

In order to disentangle the 2D and 3D structures of the flow, 
following~\cite{musacchio2017split}, 
we decompose the velocity field as ${\bm u} = {\bm  u}^{2D} + {\bm u}^{3D}$.
The 2D mode is defined in Fourier space as 
the mode $k_3=0$ of the horizontal velocity components  
$\hat{\bm  u}^{2D}(k_1,k_2) = ( \hat{u}_x(k_1,k_2), \hat{u}_y(k_1,k_2),0)$. 
This corresponds to the average in physical space along the $z$ direction. 
The 3D part is simply ${\bm u}^{3D}={\bm u}-{\bm  u}^{2D}$.
The kinetic energy $E=(1/2) \langle |{\bm u}|^2 \rangle = E^{2D}+E^{3D}$
is the sum of the two contributions 
$E^{2D}=(1/2) \langle |{\bm u}^{2D}|^2 \rangle$
$E^{3D}=(1/2) \langle |{\bm u}^{3D}|^2 \rangle$. 
Here and in the following the brackets $\langle \cdots \rangle$
indicates the spatial average 
and we notice that the mixed term has zero average.

\begin{figure}[ht]
\includegraphics[width=.49\columnwidth]{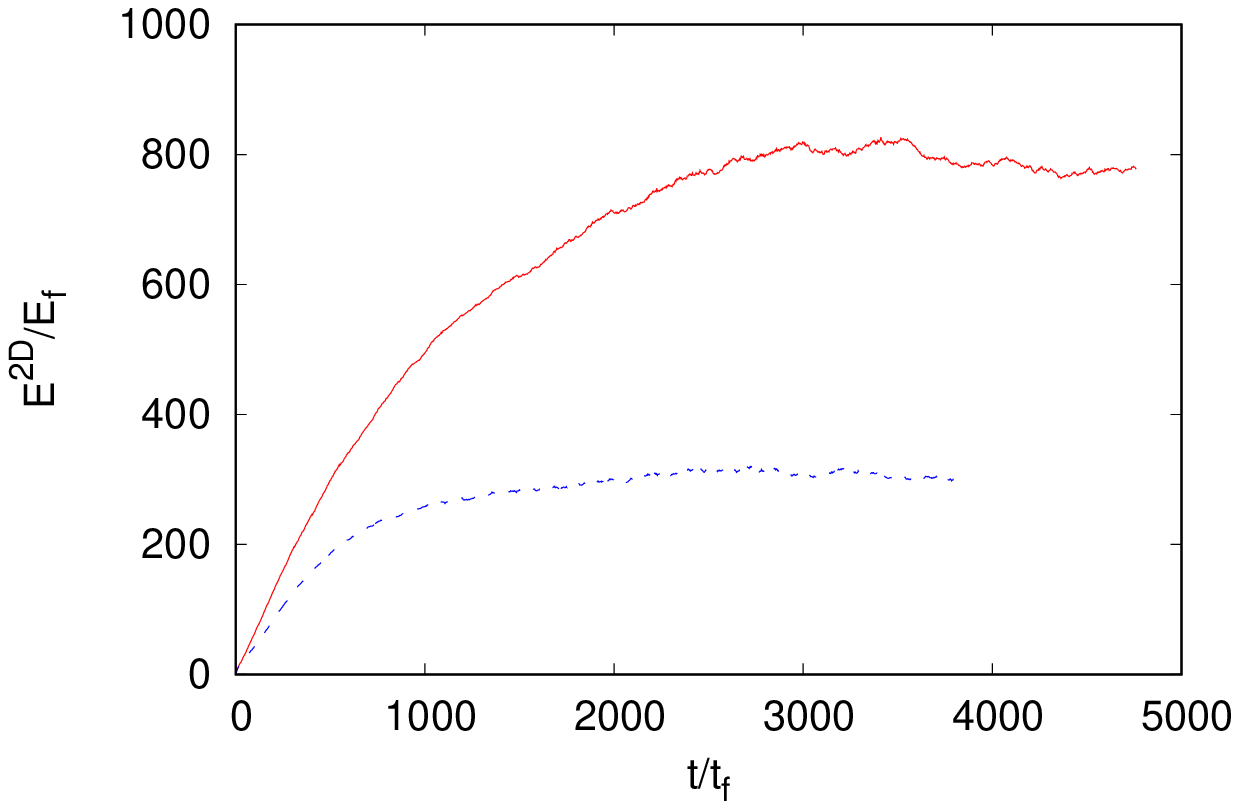}
\includegraphics[width=.49\columnwidth]{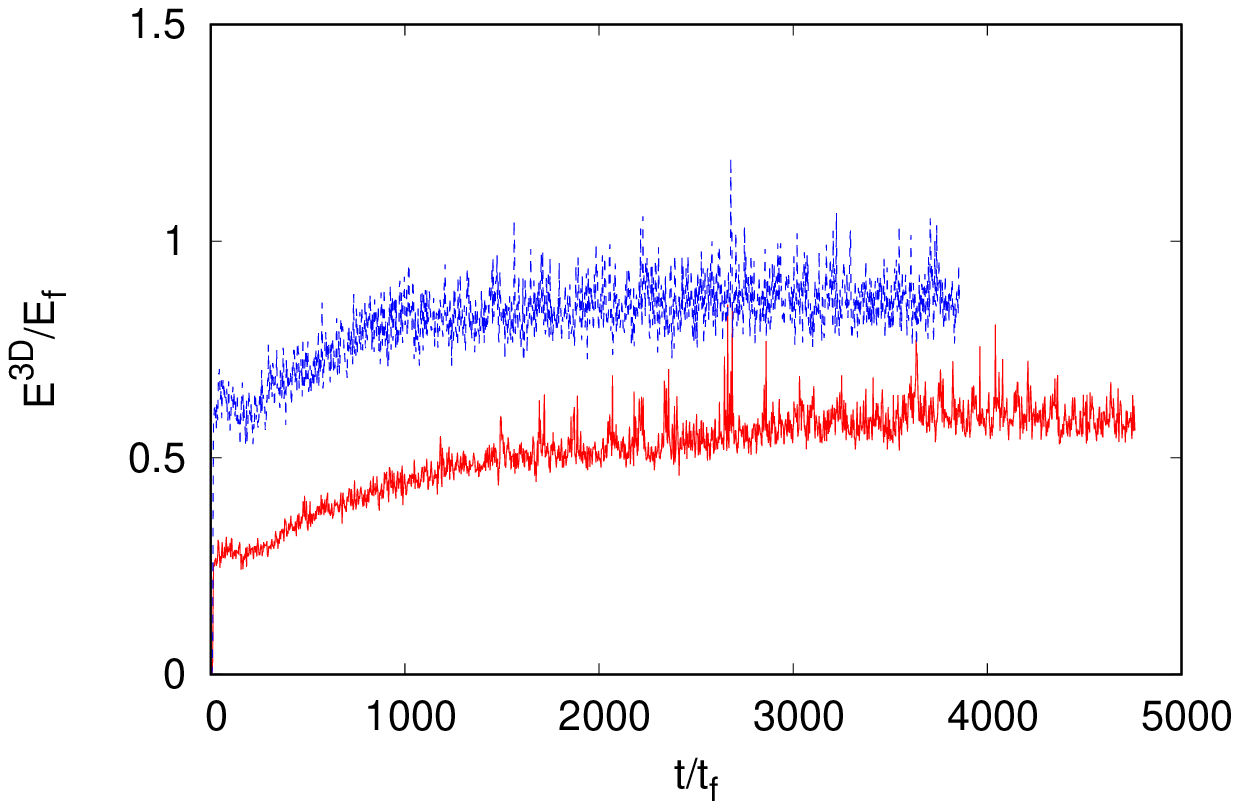}
\caption{Temporal evolution of the kinetic energy of the 2d mode
  $E^{2D}$ (left panel) and $E^{3D}$ (right panel) 
for $S=1/8$ (red, solid line) and $S=1/4$ (blue, dashed line). }
\label{fig2}
\end{figure}
The temporal evolution of $E^{2D}$ and $E^{3D}$ is shown in 
Figure~\ref{fig2}. 
For short times we observe a linear growth of
$E^{2D}$, which corresponds to the development of the inverse energy
cascade. A linear fit of the growth rate $dE^{2D}/dt$ for
$50 t_f < t < 250 t_f$ provides an
estimate of the flux $\varepsilon_{inv}$ of the inverse cascade. 
We obtain $\varepsilon_{inv} = 0.65 \varepsilon$ for $S=1/8$ 
and $\varepsilon_{inv} = 0.41 \varepsilon$ for $S=1/4$.  
The dependence of the inverse energy flux on $S$ is in agreement 
with previous numerical results \cite{smith1996crossover,celani2010turbulence}.
At later times ($t > 3000 t_f$ for $S=1/8$ and $t > 2500 t_f$ for
$S=1/4$ ) we observe the saturation of the energy 
to an almost constant value $E_c \simeq 800 E_f$ for $S=1/8$ 
and $E_c \simeq 320 E_f$ for $S=1/4$.

The energy of the 3D modes is much smaller than that of the 2D mode. 
It does not contribute significantly to the total energy. 
Its time evolution shows a first plateau at short times
($t<200 t_{f}$),
with $E^{3D} \simeq  0.28 E_f$ for $S=1/8$ and 
$E^{3D} \simeq  0.62 E_f$ for $S=1/4$, which corresponds to the 
development of the direct energy cascade.
After a slow growth, it reaches a second plateau
with $E^{3D} \simeq  0.6  E_f$ for $S=1/8$ and 
$E^{3D} \simeq  0.87  E_f$ for $S=1/4$, 
As expected, the thicker layer $S=1/4$ has smaller $E^{2D}$ 
and higher $E^{3D}$ than the thinner layer $S=1/8$. 

At variance with the case of 2D turbulence \cite{eyink1996exact},
the mechanism which arrests the growth of the condensate 
in the thin layer is not due to viscous effects at the scale of the condensate. 
A simple dimensional argument for the saturation of the energy 
due to hyperviscosity
gives an estimate of the energy of the condensate
$E^*_c \simeq \varepsilon_{inv}/ K_x^{2p} \nu_p \simeq 10^{37} E_f$, 
which is many orders of magnitude higher that the observed value of
$E_c$. This exclude the possibility that the saturation of kinetic
energy is due the hyperviscous dissipative forces at large scales. 
The dissipation spectrum, defined as 
\begin{equation}
\label{eq:spectra}
D(k_h) = 
\!\!\!\!\!
\sum_{\substack{ {\bm k} \\ k_x^2+k_y^2 = k_h^2}} 
\!\!\!\!\!
\nu_p |{\bm k}|^{2p} |\hat{\bm u}_{\bm k}|^2
\end{equation}
where $k_h = ( k_1^2+k_2^2)^{1/2}$ is the horizontal wavenumber, 
confirms that the hyperviscous dissipation is confined at high wavenumbers
and it does not affect directly the condensate 
(see Fig.{\ref{fig3}, left panel). 

\begin{figure}[ht]
\includegraphics[width=0.49\columnwidth]{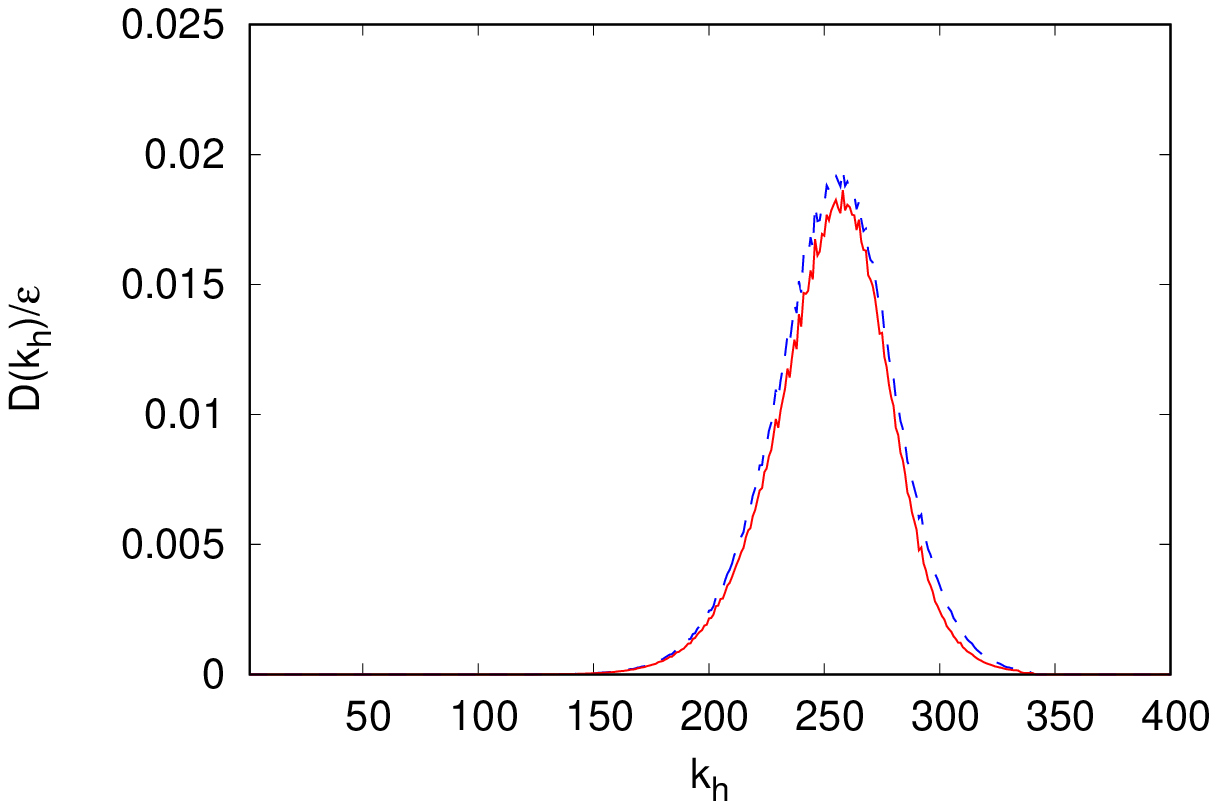}
\includegraphics[width=0.49\columnwidth]{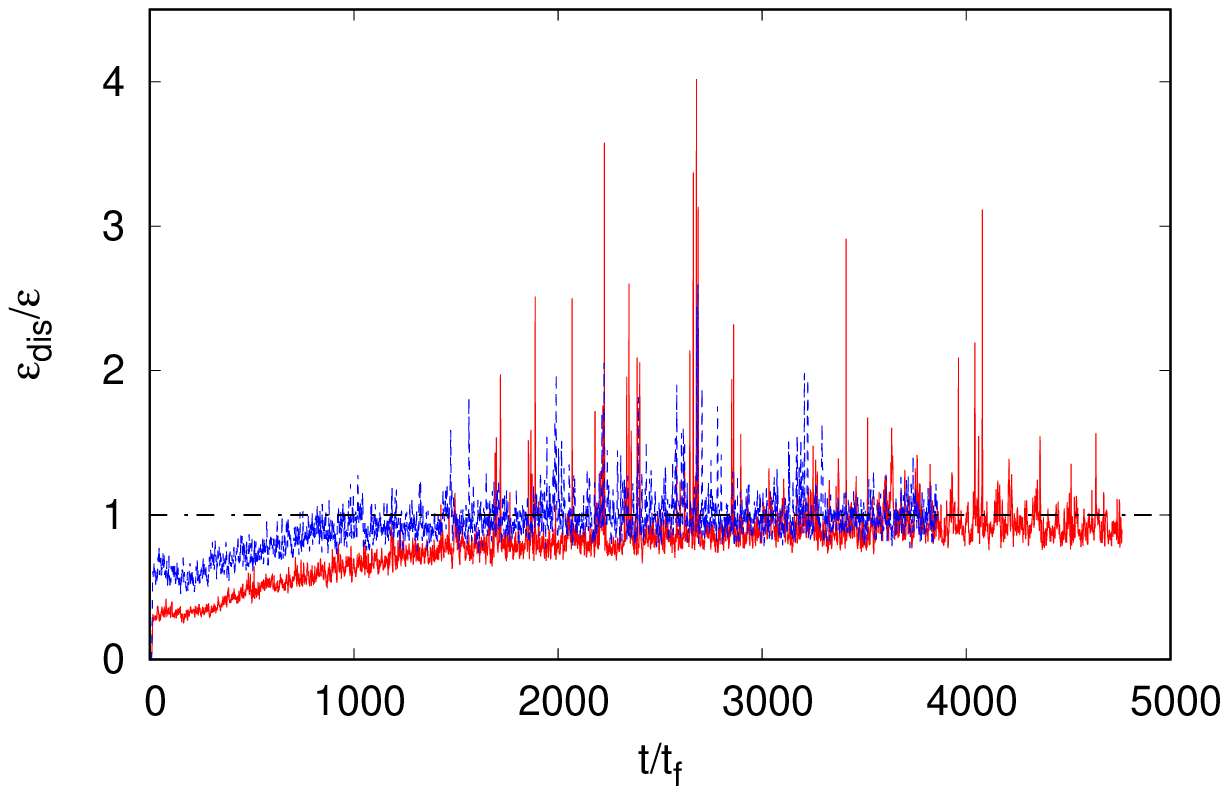}
\caption{
Left: 
Dissipation spectra $D(k_h)$ 
as a function of $k_h = ( k_x^2+k_y^2)^{1/2}$
for $S=1/8$ at time $t = 4050 t_f$ (red, solid line), 
and for 
for $S=1/4$ at time $t=2650 t_f$ (blue, dashed line). 
Right:
Temporal evolution of the energy dissipation rate 
$\varepsilon_{dis}$ for $S=1/8$ (red, solid line) and $S=1/4$ (blue,
dashed line). 
}
\label{fig3}
\end{figure}
Nonetheless, the saturation of the energy at long times 
can be achieved only if all the energy injected is dissipated at small scales,
which is the only dissipative term in the equations.
In Figure~\ref{fig3} (right panel) we show the time series of the 
(hyper-)viscous energy dissipation rate 
$\varepsilon_{dis} = \langle 2 (-1)^{p-1}\nu_p (\nabla^{2p} {\bf u})
\cdot {\bf u} \rangle$. 
After a rapid initial growth, energy dissipation displays a plateau 
for $20 t_f < t < 250 t_f$, with 
$\varepsilon_{dis}\simeq (0.32 \pm 0.02) \varepsilon$ for $S=1/8$ and 
$\varepsilon_{dis}\simeq (0.58 \pm 0.02) \varepsilon$ for $S=1/4$. 
Within the scenario of the split-energy
cascade~\cite{musacchio2017split}, 
the dissipation rate is equal to the flux of the direct energy cascade
$\varepsilon_{dir} = \varepsilon - \varepsilon_{inv}$. 
The measured values of $\varepsilon_{dis}$ are consistent with this picture. 
At long times, the large-scale 2D condensate interact directly with 
the small-scales 3D flow, transferring its energy toward the 3D modes 
with $|{\bm k}| >k_z$ (at the rate $\varepsilon_{inv}$) where it is 
transported by the direct cascade to the dissipative scales. 
In this regime, which corresponds to the saturation
of kinetic energy in Fig.~\ref{fig2}, we therefore observe that 
$\varepsilon_{dis}$ tends to values close to the energy
input rate $\varepsilon$.
We notice that for $t > 1000 t_f$ the dissipation becomes
extremely intermittent in time.
This coupling between the large scale condensate and small scale 3D motion 
will be investigated in details in the following. 

In order to make quantitative predictions of the saturation energy of
the condensate, we suppose that it is possible to model the 3D dynamics
at scales $\ell < L_z$ and its interactions with the condensate by
means of an effective eddy viscosity $\nu_{eddy}$, which is 
much larger than the molecular (hyper) viscosity. 
Further, we assume that that the condensate reaches a steady state when the flux of
the inverse cascade which feeds it, is balanced by the effect of the
eddy viscosity at the scale $L_x$, that is, 
$\varepsilon_{inv} \simeq \nu_{eddy} E_c/ L_x^2$. 
Using the simple dimensional estimate for the eddy viscosity 
$\nu_{eddy} \simeq E^{3D} t_f$ we obtain 
scaling predictions for the energy of the condensate $E_c$
and the time required to form it $t_c \simeq E_c / \varepsilon_{inv}$:   
\begin{equation}  
\label{eq:econd}
\frac{E_c}{E_f} \simeq
\frac{E_f}{E^{3D}} 
\left(\frac{L_x}{L_f}\right)^2
\frac{\varepsilon_{inv}}{\varepsilon}
; \;\;\;\;  
\frac{t_c}{t_f} \simeq 
\frac{E_f}{E^{3D}} 
\left(\frac{L_x}{L_f}\right)^2
\end{equation}
Using the values of $\varepsilon_{inv}$ and $E^{3D}$ 
measured in the simulations, we get quantitative estimates for $E_c$
and $t_c$. Rescaling the time and energy with the predictions~(\ref{eq:econd}), 
we observe a good collapse of the temporal evolution of the
kinetic energy (see Fig.~\ref{fig4} left panel). 
The validation of the predictions~(\ref{eq:econd})
would require a large set of simulations with different aspect
ratios. This is left for future investigations. 

In the asymptotic limit of infinite $Re$ and small $S$, 
it is possible to derive theoretical scaling predictions for $E_c$
and $t_c$ as a function of the aspect ratios 
$S=L_z/L_f$ and $r=L_x/L_z$ only.
According to the phenomenology described
in~\cite{musacchio2017split}, three cascade processes 
take place in the turbulent layer. 
At large scales ($\ell > L_f$) there is a 2D 
inverse energy cascade with flux $\varepsilon_{inv}$. 
At intermediate scales ($L_f > \ell > L_z$) 
the enstrophy production is negligible, and 
a 2D direct enstrophy cascade with flux $\beta$ is observed. 
At small scales ($ \ell < L_z$) the flow becomes 3D 
and displays a direct energy cascade with flux 
$\varepsilon_{dir}$, which is assumed to be proportional
to the residual flux of energy 
carried by the enstrophy cascade at the scale $L_z$, that is,  
$\varepsilon_{dir} \propto \beta L_z^2$~\cite{musacchio2017split}. 
Recalling that $\varepsilon_{inv} = \varepsilon - \varepsilon_{dir}$, 
and that $\varepsilon_{inv} =0$ for $S$ larger than the critical aspect
ratio $S^* =L^*_z/L_f$~\cite{celani2010turbulence},
one gets $\varepsilon_{inv}/\varepsilon \propto 1-(S/S^*)^2$.  
Similarly, one can estimate the energy of the 3D mode as $E^{3D} \simeq
u_z^2$, where $u_z$ is the typical intensity of the velocity at the scale $L_z$. 
From the scaling of the direct enstrophy cascade one has
$u_z \simeq \beta^{1/3} L_z$, and hence 
$E^{3D} \simeq \beta^{2/3} L_z^2 \simeq E_f (L_z/L_f)^2$. 
Inserting these dimensional estimates in Eq.~\ref{eq:econd} one obtains
the asymptotic scaling predictions: 
\begin{equation}  
\label{eq:econd_scaling}
\frac{E_c}{E_f} \simeq
\left(\frac{L_x}{L_z}\right)^2
\left[ 1- \left(\frac{L_z}{L^*_z}\right)^2 \right]
 \;\;\;\;  
\frac{t_c}{t_f} \simeq 
\left(\frac{L_x}{L_z}\right)^2. 
\end{equation}
The asymptotic scaling requires a very high Reynolds number to be 
verified, which cannot be achieved in fully resolved simulations
but has been observed in simplified dynamical model of turbulence
\cite{boffetta2011shell}.

Following Ref.~\cite{laurie2014universal}, we have computed 
the mean 2D vorticity field of the condensate, 
as the temporal average in the stationary regime of the fields $\omega_z$, 
which have been previously averaged in the vertical direction $z$ and
then centered on the position of the 
center of mass of the vorticity. 
The radial vorticity profile of the condensate $\Omega(r)$ 
(shown in Fig.\ref{fig4}, right panel) 
is characterized by a vortex core (for $r < R_c$) in which the vorticity
is almost constant, as in solid body rotation.  
Rescaling the radial distance with a dimensional estimate for 
the radius of the core $R_c \simeq
\sqrt{E/Z}$, where $Z=(1/2) \langle |{\bm \omega}|^2 \rangle$ is the enstrophy,
we observe a good collapse of the core (see inset of Fig.\ref{fig4}, right panel). 
The radius $R_c$ represents the scale at which the
centripetal force of the vortex is balanced by the inertial forces of
turbulence.

Outside the core
we observe for the case $S=1/4$, a power law
behaviour for the vorticity profile $\Omega(r) \sim r^{-1} $, 
which is suggestive of a similar observation in 2D
turbulence~\cite{laurie2014universal}. 
However, in~\cite{laurie2014universal} 
this scaling is derived from the
balance with a friction force which is not
present in our 3D case. Different predictions has been derived for a
2D viscous condensate~\cite{kolokolov2015profile}. 
In both cases~\cite{laurie2014universal,kolokolov2015profile} 
the 2D scalings are expected for scales $r> L_f$ 
while in our simulation the scaling range is $R_c < r < L_f$.
Further investigations are required to achieve a better understanding
of the profile of the 3D condensate and its relations with the 2D
case. 

\begin{figure}[ht]
\includegraphics[width=0.49\columnwidth]{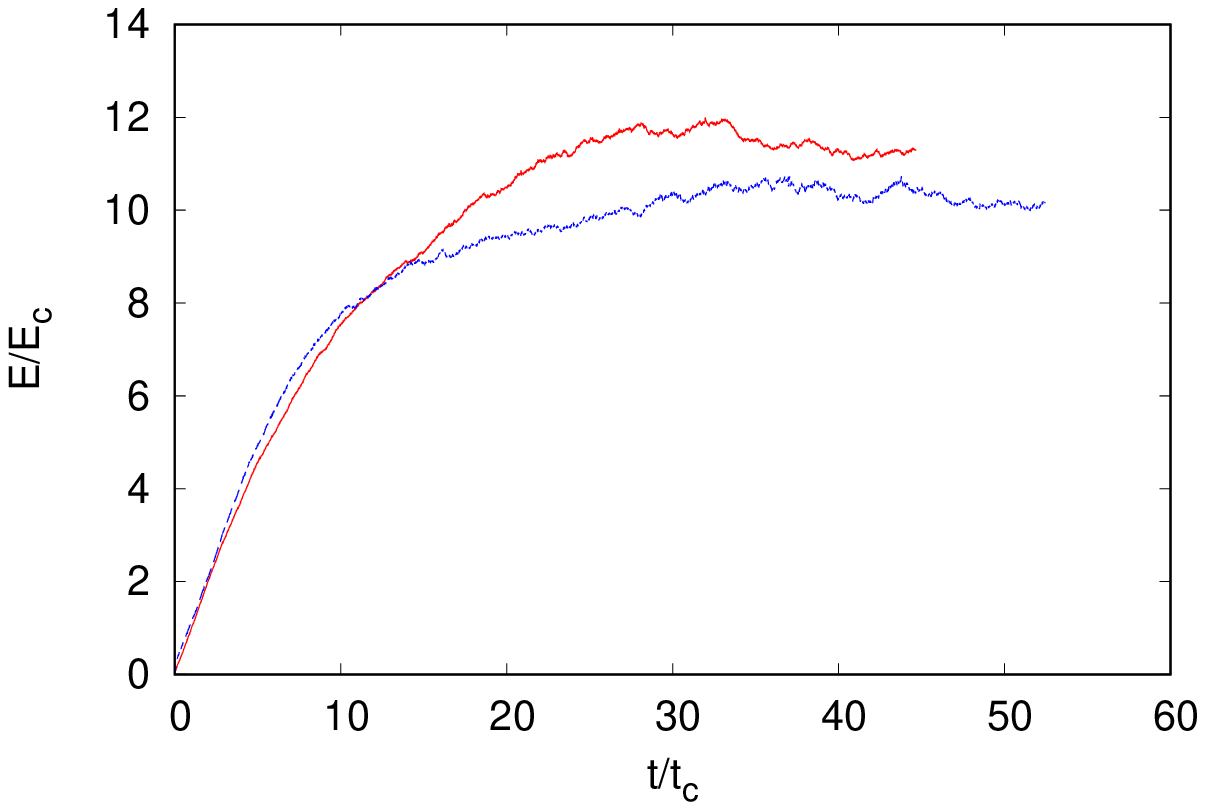}
\includegraphics[width=0.49\columnwidth]{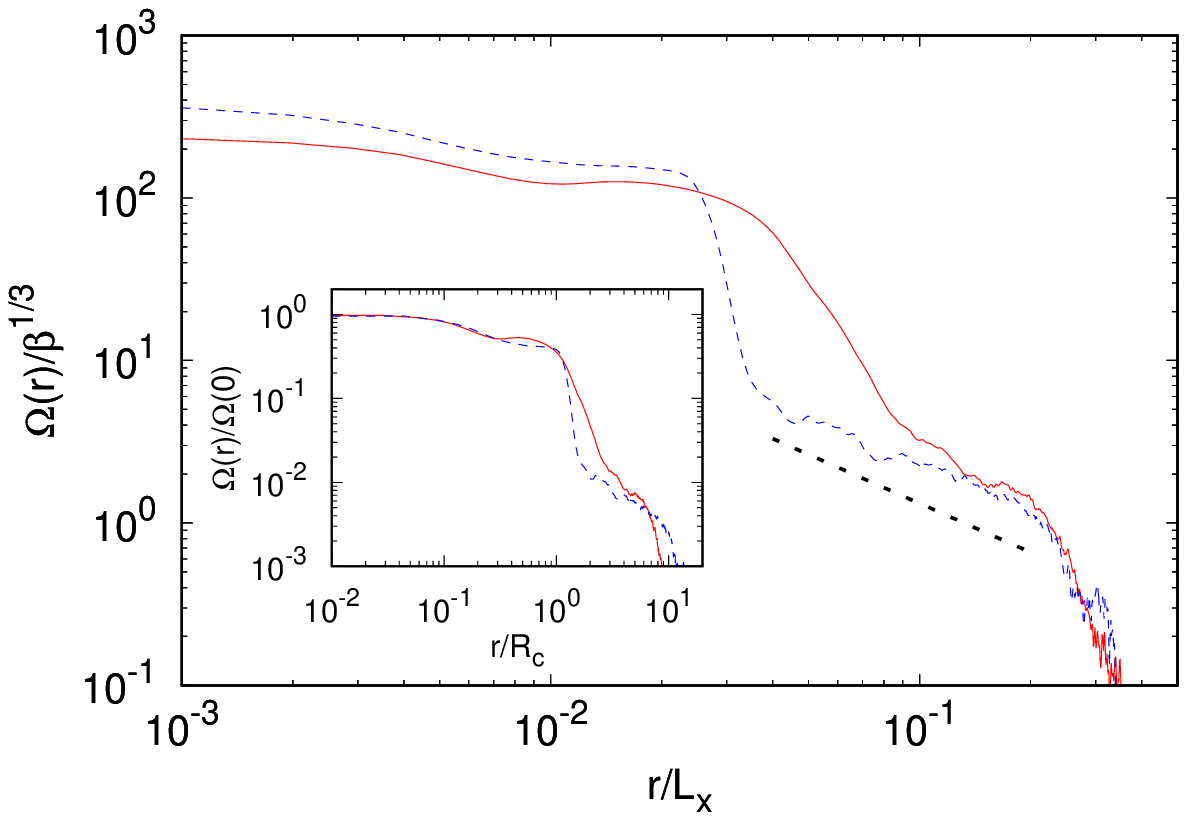}
\caption{
Left: 
Temporal evolution of the kinetic energy 
rescaled with $E_c$ and $t_c$ estimated by ~(\ref{eq:econd}).
for $S=1/8$ (red, solid line) and $S=1/4$ (blue, dashed line). 
Right:
Radial profile of the mean vorticy of the condensate
 $\Omega(r)$ for $S=1/8$ (red, solid line) and $S=1/4$ (blue, dashed
 line). The scaling behavior $\Omega(r) \sim r^{-1}$ is represented
 by the black dotted line.
}
\label{fig4}
\end{figure}

In Figure~\ref{fig5} we show the 2D energy spectra 
for the case $S=1/4$,  
defined as 
\begin{equation}
\label{eq:spectra}
E(k_h) = \frac{1}{2} 
\!\!\!\!\!
\sum_{\substack{ {\bm k} \\ k_x^2+k_y^2 = k_h^2}} 
\!\!\!\!\!
|\hat{\bm u}_{\bm k}|^2
\end{equation}
where $k_h = ( k_1^2+k_2^2)^{1/2}$ is the horizontal wavenumber. 
At short time the spectra shows the development of the inverse energy
cascade for $k_h > k_f$ with a $-5/3$ spectral slope.  
At late times, the spectrum of the steady condensed state contains
much more energy than the spectrum of the inverse cascade 
in a broad wavenumber range ($k_h > k_z$). 
In the range $k_h > k_f$ the spectral slope is close to $-2$. 
A similar spectral behavior is observed for the case $S=1/8$ (not shown). 
\begin{figure}[h!]
\centering
\includegraphics[width=0.49\columnwidth]{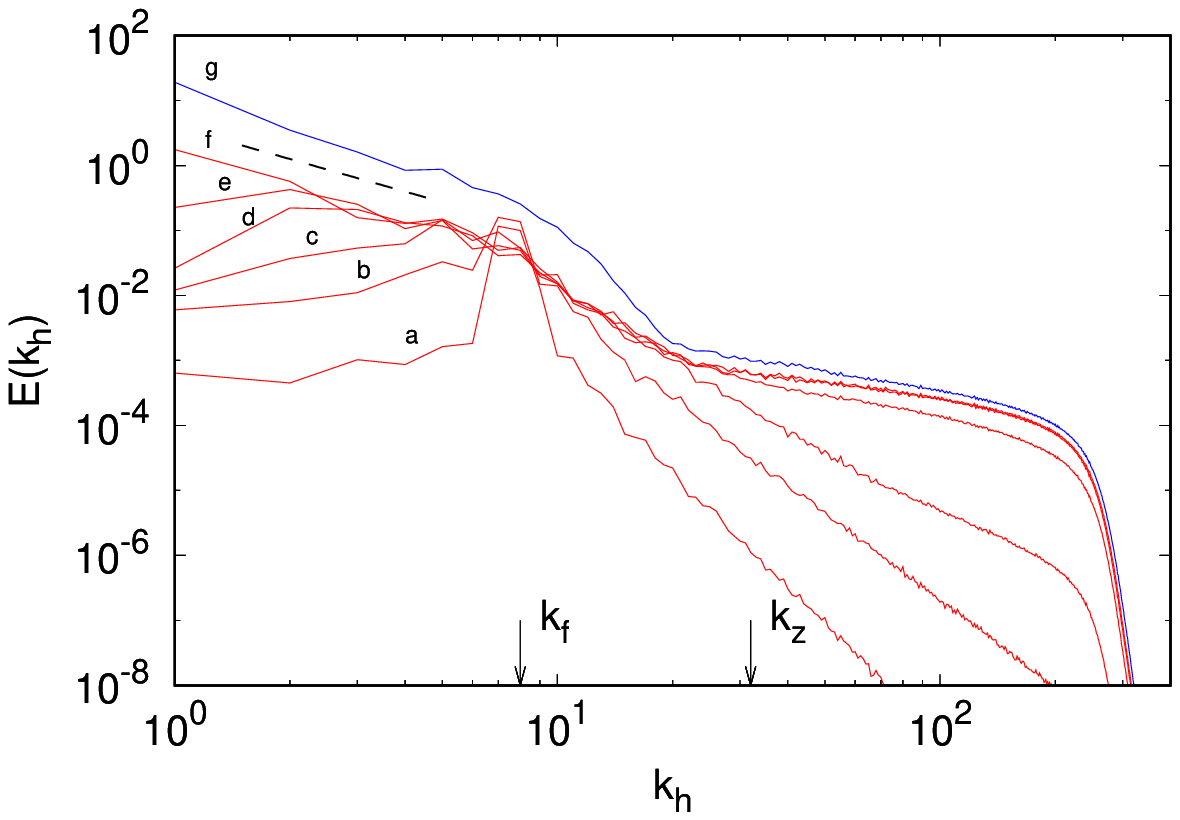}
\includegraphics[width=0.49\columnwidth]{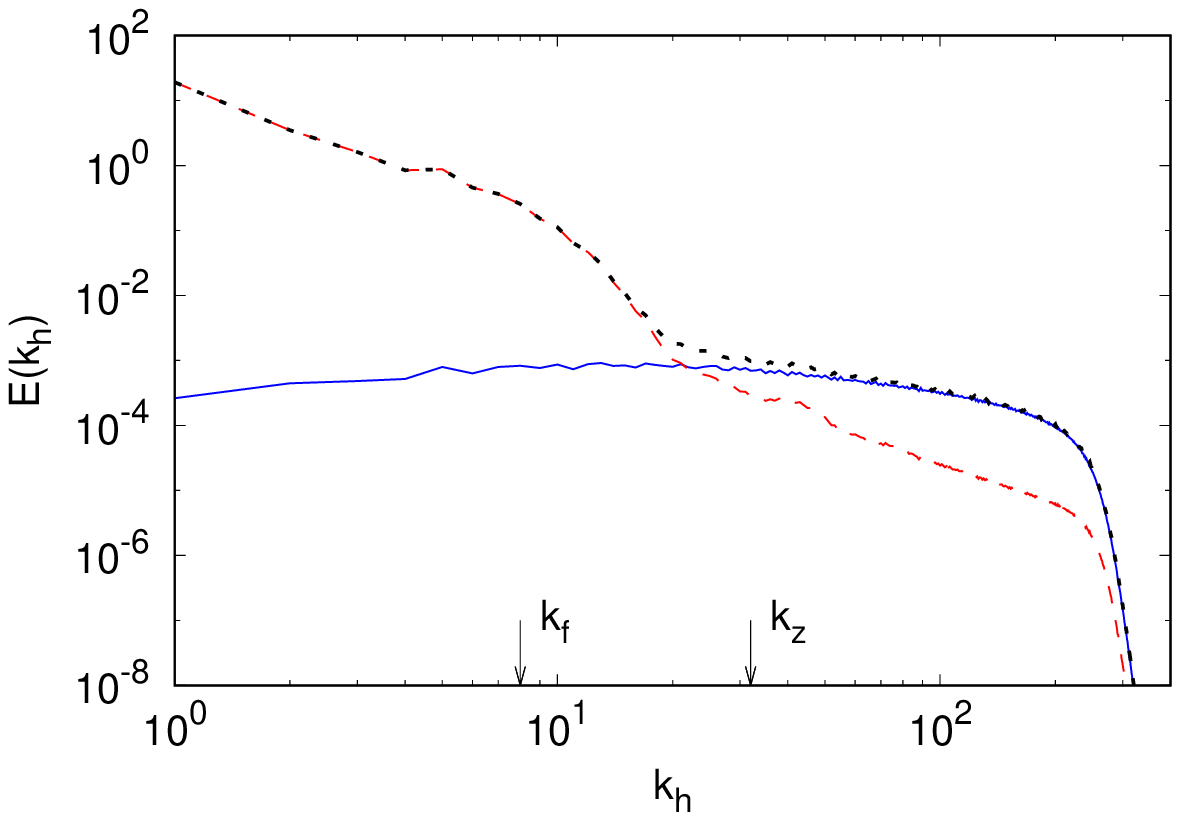}
\caption{
Left: 
2D Energy spectra $E(k_h)$ 
as a function of $k_h = ( k_x^2+k_y^2)^{1/2}$
for $S=1/4$ at times $t/t_f =$ 
$2.4$ (a), 
$4.8$ (b), 
$7.1$ (c), 
$12$ (d), 
$24$ (e), 
$71$ (f), 
and
$2650$ (g).
The black dashed line represent the scaling $k^{-5./3.}$. 
Right: 
2D Energy spectrum $E(k_h)$ 
as a function of $k_h = ( k_x^2+k_y^2)^{1/2}$
of the 
2D mode ${\bm u}^{2D}$ (red, dashed line), 
3D mode ${\bm u}^{3D}$ (blue, solid line) and 
total velocity field ${\bm u}$  (black, dotted line) 
as a function of the horizontal wavenumber $k_h$.  
}
\label{fig5}
\end{figure}

In order to highlight the different contribution of the 2D and 3D modes to the
spectrum of the condensate, we show in Figure~\ref{fig5} the 2D
energy spectra of the fields $ {\bm  u}^{2D}$ and ${\bm u}^{3D}$.
The energy of the 2D mode is dominant for $k_h > k_z$, which 
confirms the two-dimensional nature of the condensed state. 
Conversely, the 3D mode becomes dominant at small scales $k_h > k_z$. 

The spectral energy flux, defined as
\begin{equation}
\Pi(k) = - \frac{1}{2} 
\!\!\!\!\!
\sum_{
\substack{
\bm k, \bm p, \bm q \\
|{\bm k}| \le k \\
{\bm k} + {\bm p} + {\bm q} = 0
}}
\!\!\!\!\!
\hat{\bm u}^*_{\bm k} 
\cdot \left(
i {\bm k} \cdot \hat{\bm u}_{\bm p}
\right)
\hat{\bm u}_{\bm q}
+ c.c.
\label{eq:flux} 
\end{equation}
gives further informations concerning the mechanisms of the formation
and saturation of the condensate. 
In the early stage in which the condensate grows, ($t=25t_f$ in Fig.~\ref{fig6})
the energy flux shows clearly the splitting of the energy cascade 
(as in~\cite{celani2010turbulence, musacchio2017split}).  
A fraction of the energy is transported toward small wavenumbers
$k<k_f$ with a negative flux $\varepsilon_{inv}$, while the remnant
energy perform a direct cascade toward large $k$ with flux $\varepsilon_{dir}$
In the late stage ($t=2650t_f$ in Fig.\ref{fig6})
when the condensate has reached a steady state, the average flux is zero
for $k < k_f$ and is equal to the energy input $\varepsilon$ for $k>
k_f$. Nonetheless, the flux of the 2D mode 
transported by the 2D velocity $\Pi^{2D}(k)$ (defined as in
Eq.~(\ref{eq:flux}) but restricting the fields to ${\bm u}^{2D}$)
reveals that also at late times, there is a negative flux of 2D energy
at $k < k_f$. Therefore the condensate is still fed 
by a 2D inverse energy cascade. 
The negative energy flux, which proceeds from the forcing scale to
the scale of the condensate, is exactly
balanced by an opposite energy transfer from the condensate
to the small scale 3D flow. 

\begin{figure}[h!]
\centering
\includegraphics[width=0.49\columnwidth]{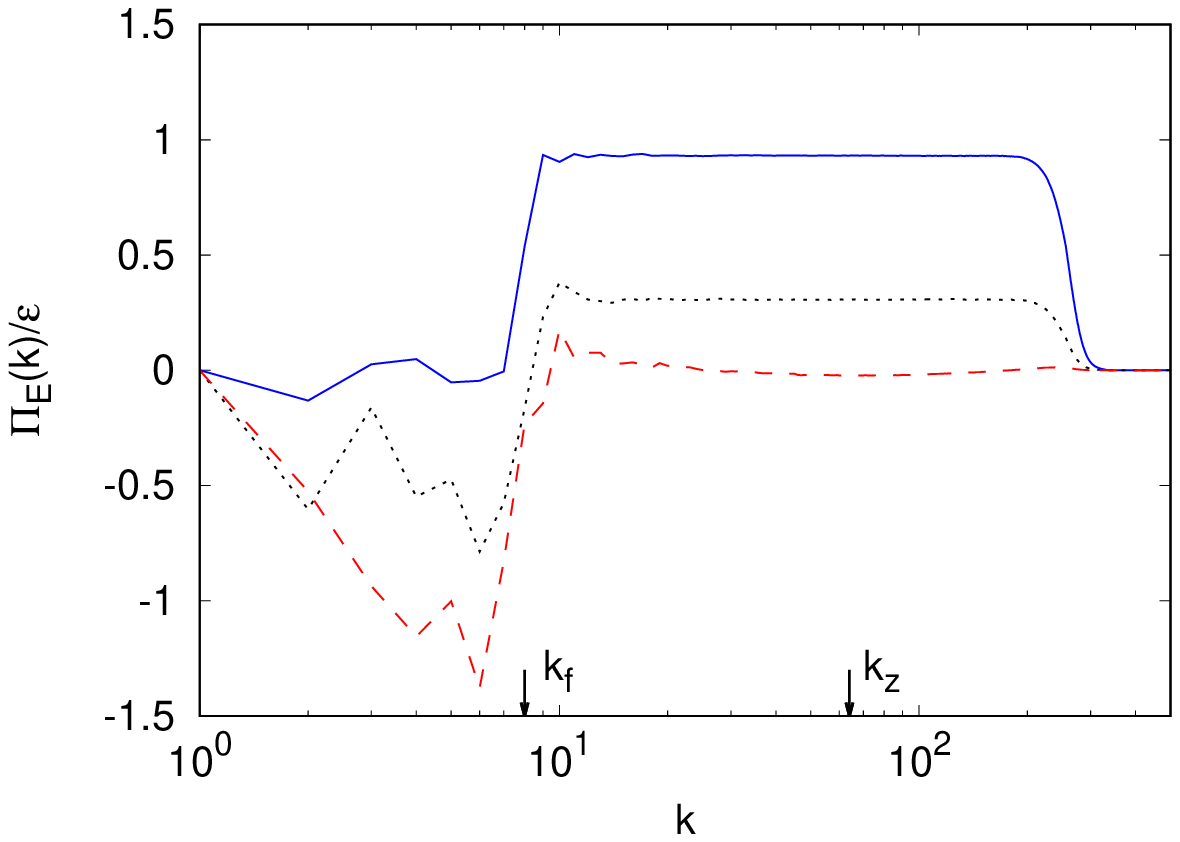}
\includegraphics[width=0.49\columnwidth]{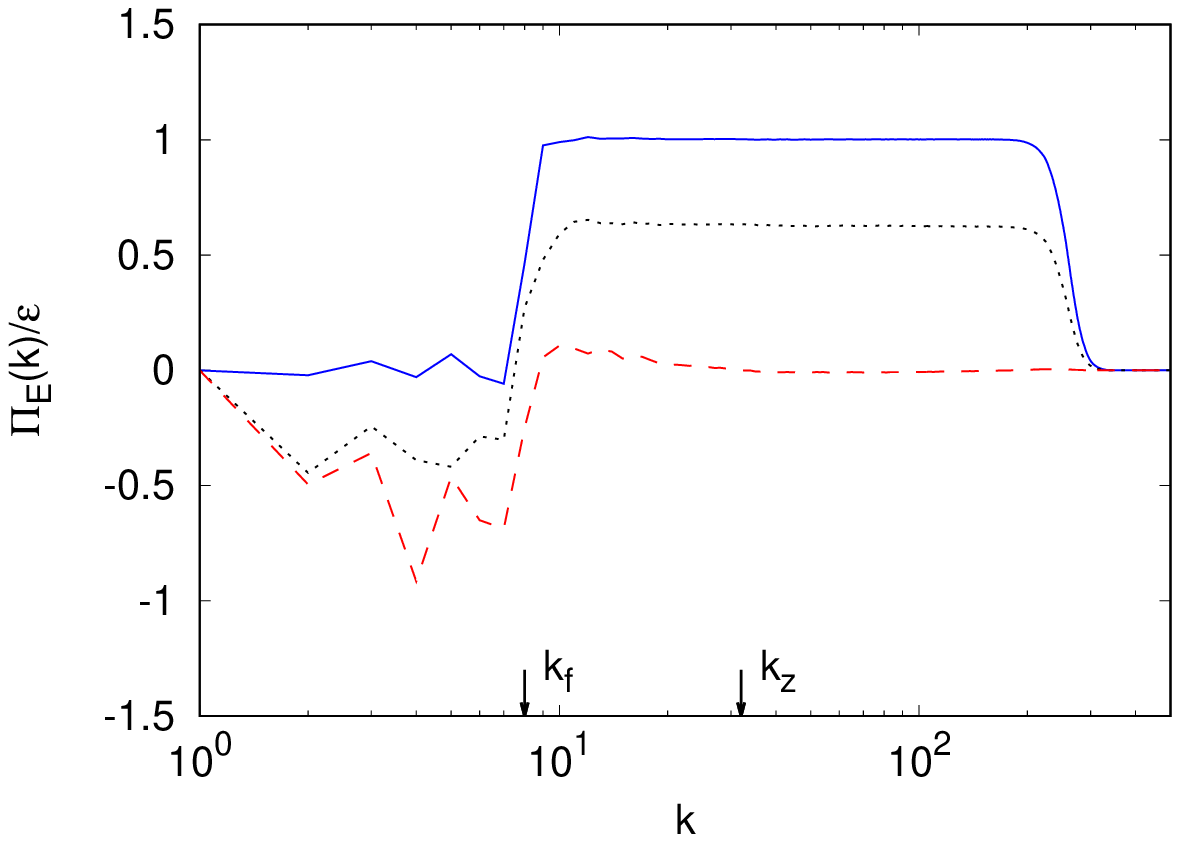}
\caption{
Left: Spectral energy flux $\Pi(k)$ for $S=1/8$ 
at time 
$t=25 t_f$ (black, dotted line)
$t=4050 t_f$ (blue, solid line), 
and 2D energy flux  $\Pi^{2D}(k)$ at $t = 4050 t_f$
(red, dashed line). 
Right: Spectral energy flux $\Pi(k)$ for $S=1/4$ 
at time 
$t=25 t_f$ (black, dotted line)
$t=2650 t_f$ (blue, solid line), 
and 2D energy flux  $\Pi^{2D}(k)$ at $t = 2650 t_f$
(red, dashed line). 
}
\label{fig6}
\end{figure}

To investigate the interactions between the 2D and 3D modes at
different scales, we partition the Fourier space in non-overlapping
spherical shells $n \Delta K \le |k|< (n+1) \Delta K$, 
labelled with $K=1+n\Delta K$, with $\Delta K=4$. 
Then, we decompose the velocity field as 
${\bm u} = \sum_K {\bm u}_K$, 
where ${\bm u}_K$ is the velocity field filtered in the shell $K$.  
Following~\cite{alexakis2005imprint} we define the rate of energy
transfer $T(K,Q)$ from the shell $Q$ to the shell $K$ as: 
\begin{equation}
T(K,Q) = - 
\sum_{P}
\int d{\bm x}^3
{\bm u}_K
\cdot \left(
{\bm u}_P\cdot {\bm \nabla} \right)
{\bm u}_{Q}
\label{eq:transfer} 
\end{equation}
Similarly, we define the 2D transfer $T^{2D}(K,Q)$ in terms of the
filtered 2D modes ${\bm u}_K^{2D}$. 
In Figure~\ref{fig7} we plot the spectral energy transfer toward 
the shell $K=1$ which contains the condensate. 
In order to reduce the statistical fluctuations, 
we have averaged $T(K,Q)$ and $T^{2D}(K,Q)$ for times larger than the time $t_s$ 
required to reach the statistically steady state. 
In both cases $S=1/4$ and $S=1/8$ we observe that the shells with 
$Q < k_z$ give positive contributions to the condensate. In this range
of wavenumbers, the 2D transfer  $T^{2D}(K,Q)$ coincides with the total
transfer $T(K,Q)$, showing that the condensate is fed 
solely by the interactions between large-scale 2D modes. 
Conversely, the modes $Q \ge k_z$ take away energy from the
condensate, but their negative contributions to $T(K,Q)$ 
are canceled out in $T^{2D}(K,Q)$. 
This demonstrates that the condensate reaches a statistically steady
state because of interactions with the small-scale 3D modes which subtract energy from it. 
The non-local nature of these interactions is evident, 
because they occurs between the shell $K=1$ and the modes $Q \ge k_z \gg 1$. 

\begin{figure}[h!]
\centering
\includegraphics[width=0.49\columnwidth]{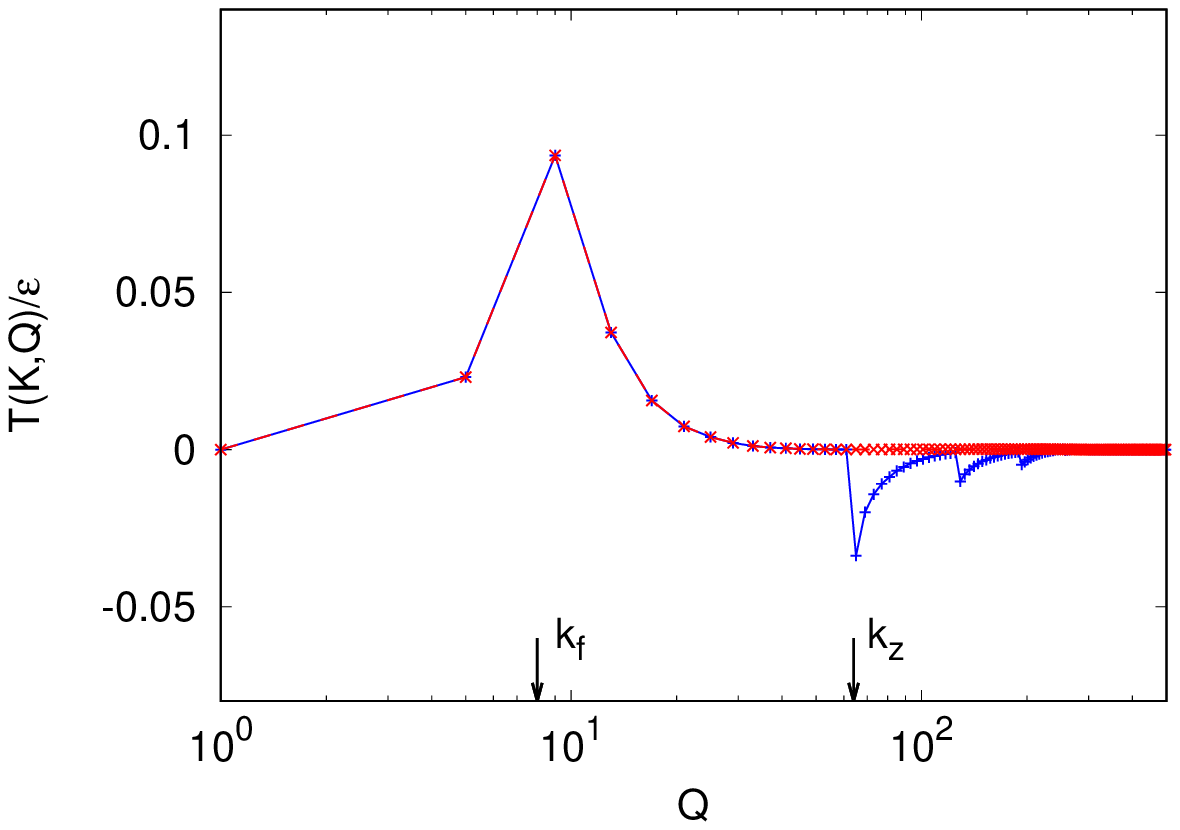}
\includegraphics[width=0.49\columnwidth]{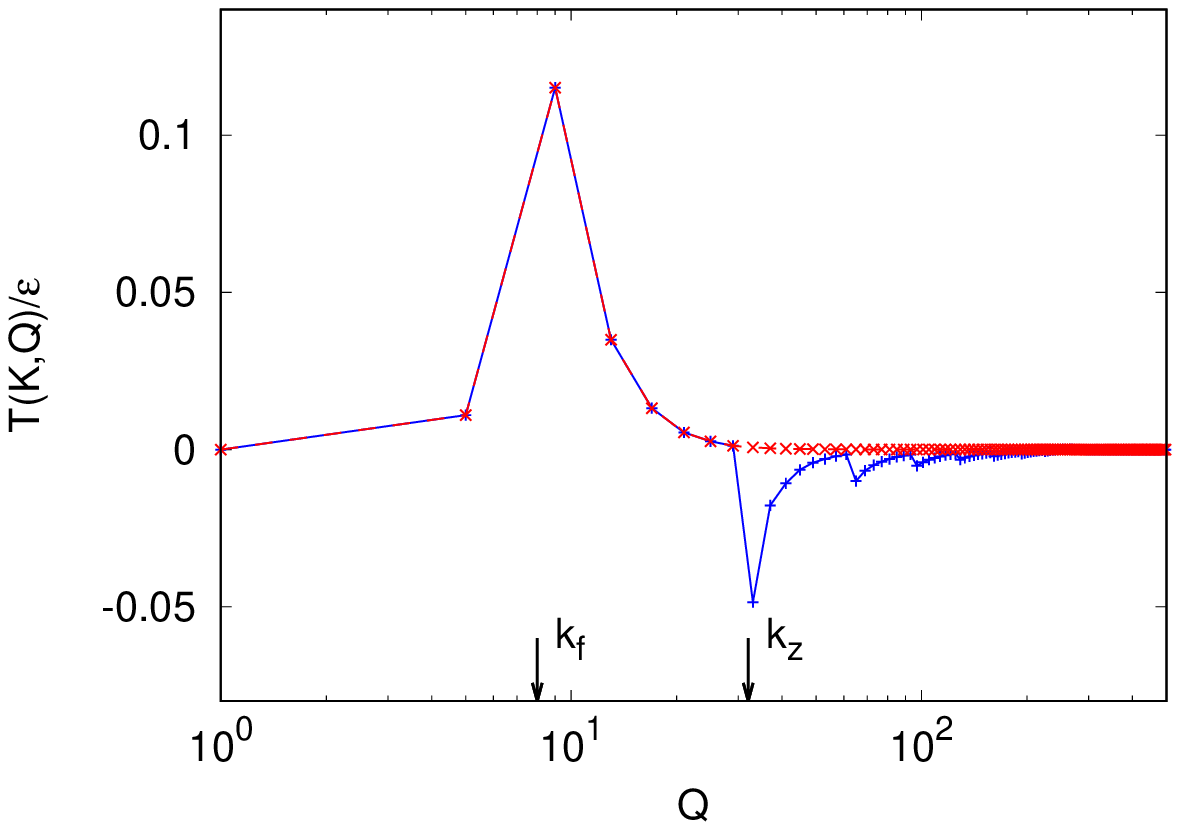}
\caption{
Spectral energy transfer $T(K,Q)$ (blue, crosses), 
and 2D spectral energy transfer  $T^{2D}(K,Q)$ (red, times)
for $K=1$, averaged in time for $t > t_s$. 
Left panel: $S=1/8$,  $t_s=3000 t_f$. 
Right panel: $S=1/4$, $t_s=2500 t_f$. 
}
\label{fig7}
\end{figure}

In conclusion, we have shown that by confining a the turbulent flow,
forced at small scale $L_f$, in a thin fluid layer, it is possible to 
observe the formation of a
statistically steady condensed state, which has the form of a quasi
two-dimensional dipole. 
By means of direct numerical simulations we have analyzed the
temporal evolution of the kinetic energy and its spectral
distribution, showing that the condensate is composed mainly by the 
the 2D mode. 
We have also demonstrated that the saturation of the energy of the condensate
is due to the balance of two processes: an inverse cascade of 2D
energy, which proceeds from the forcing scale toward the large scales
$\ell > L_f$, and a direct energy transfer from the
condensate toward the 3D turbulent flow at small scales $\ell <
L_z$.  The latter process is similar to the viscous dissipative
process which arrests the cascade in a purely 2D flow, but the
role of the viscosity is here replaced by the eddy viscosity of the 3D
flow. 

It is worth remarking that, because of the dissipative anomaly,  
the eddy viscosity does not vanish in the limit $\nu
\to 0$. This guarantees the saturation of the energy to a finite 
value also in the limit of $Re \to \infty$. 
This result is of particular interest for geophysical applications, 
in which the vertical scale $L_z$ of the fluid layers is much larger
than the viscous scales. 
In view of possible application of our findings to more realistic
geophysical situations, it would be extremely interesting to investigate 
the problem with different boundary conditions and 
how the presence of the Coriolis force and of a stable stratification 
of density, affects the formation of the condensate and its saturation. 

\begin{acknowledgments}
G.B. acknowledges financial support by the project CSTO162330 
{\it Extreme Events in Turbulent Convection} and from
the {\it Departments of Excellence} grant (MIUR).
HPC center CINECA is gratefully acknowledged for computing resources.
\end{acknowledgments}

\bibliography{biblio}

\end{document}